\documentclass[showpacs,amsmath,amssymb,twocolumn,pra,superscriptaddress]{revtex4-2}

\usepackage{graphicx}
\usepackage{dcolumn}
\usepackage{bm}

\usepackage{amsmath}
\usepackage{physics}
\usepackage[justification=Justified,singlelinecheck=false,labelfont=bf]{caption}  
\usepackage{subcaption}
\usepackage{float}
\usepackage{braket}
\usepackage{comment}
\usepackage{color}
\setcitestyle{super} 
\usepackage{braket}
\usepackage{hyperref}

\newcommand{\orcidicon}{\includegraphics[width=8pt]{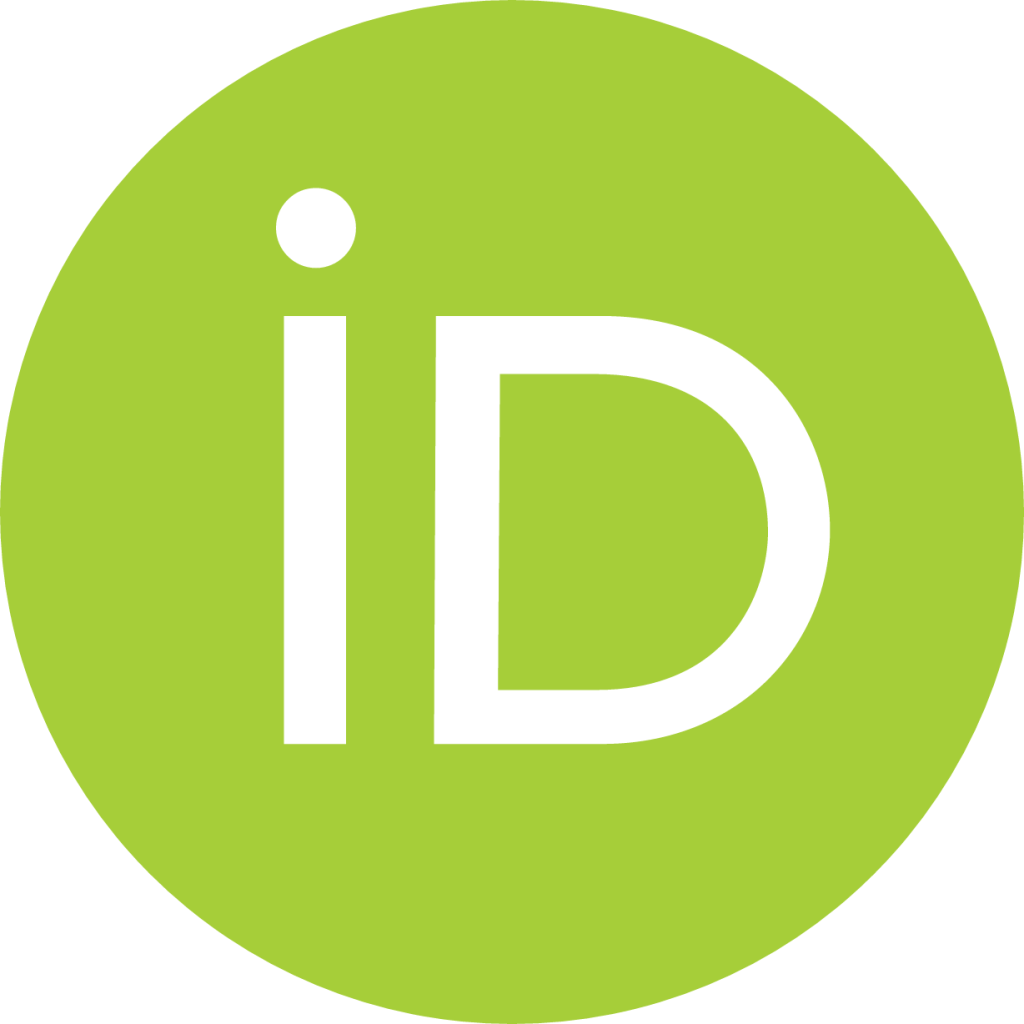}} 

\begin{document}

\title{Entanglement-Enhanced Neyman-Pearson Target Detection}%

\author{William Ward \href{https://orcid.org/0009-0002-0139-6935}{\orcidicon}}
\email{wardw@umich.edu}
\affiliation{Applied Physics Program, University of Michigan, Ann Arbor, MI 48104, USA}
\affiliation{%
 Department of Electrical Engineering and Computer Science, University of Michigan, Ann Arbor, MI 48104, USA}
\cite{} \author{Abdulkarim Hariri \href{https://orcid.org/0009-0006-8654-4616}{\orcidicon}}%
 \email{hariri@umich.edu}
 \affiliation{%
 Department of Electrical Engineering and Computer Science, University of Michigan, Ann Arbor, MI 48104, USA}
\author{Zheshen Zhang \href{https://orcid.org/0000-0002-8668-8162}{\orcidicon}}%
 \email{zszh@umich.edu}
\affiliation{%
 Department of Electrical Engineering and Computer Science, University of Michigan, Ann Arbor, MI 48104, USA}
\affiliation{Applied Physics Program, University of Michigan, Ann Arbor, MI 48104, USA}

\date{\today} 

\begin{abstract}
Quantum illumination (QI) provides entanglement-enabled target-detection enhancement, despite operating in an entanglement-breaking environment. Existing experimental studies of QI have utilized a Bayesian approach, assuming that the target is equally likely to be present or absent before detection, to demonstrate an advantage over classical target detection. However, such a premise breaks down in practical operational scenarios in which the prior probability is unknown, thereby hindering QI's applicability to real-world target-detection scenarios. In this work, we adopt the Neyman-Pearson criterion in lieu of the error probability for equally likely target absence or presence as our figure of merit for QI. We demonstrate an unconditional quantum advantage over the optimal classical-illumination protocol as benchmarked by the receiver operating characteristic, which examines detection probability versus false-alarm probability without resorting to known prior probabilities. Our work represents a critical advancement in adapting quantum-enhanced sensing to practical operational settings.
\end{abstract}

\maketitle 
\begin{figure*}[ht]
    \fbox{
    \begin{subfigure}[t]{0.47\textwidth}
        \caption{\footnotesize}
        \includegraphics[width=\linewidth]{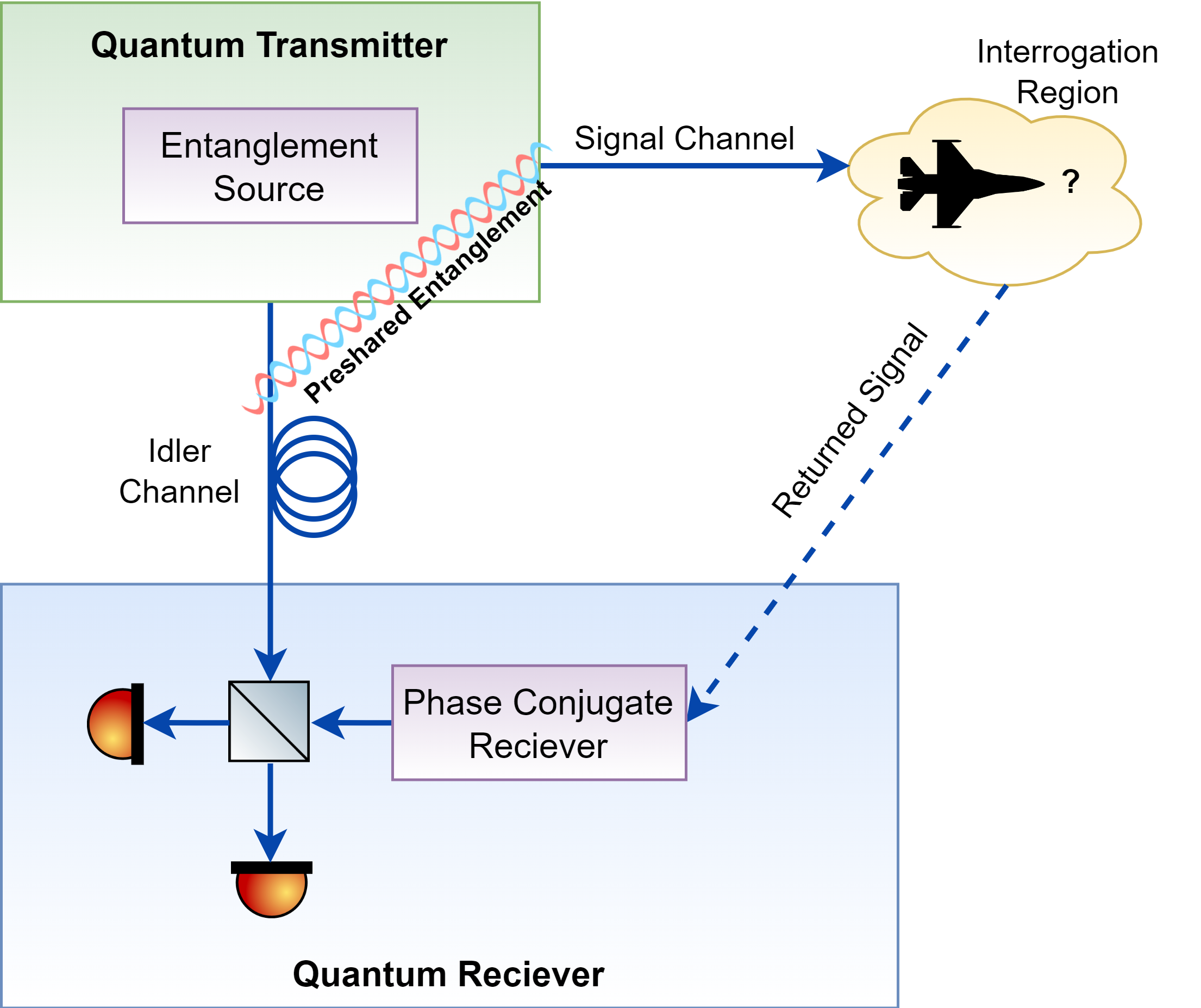}
        \label{subfig:concept_main}
    \end{subfigure}\hfill
    
    \begin{subfigure}[t]{.53\textwidth}
        \caption{\footnotesize}
        \includegraphics[width=\linewidth]{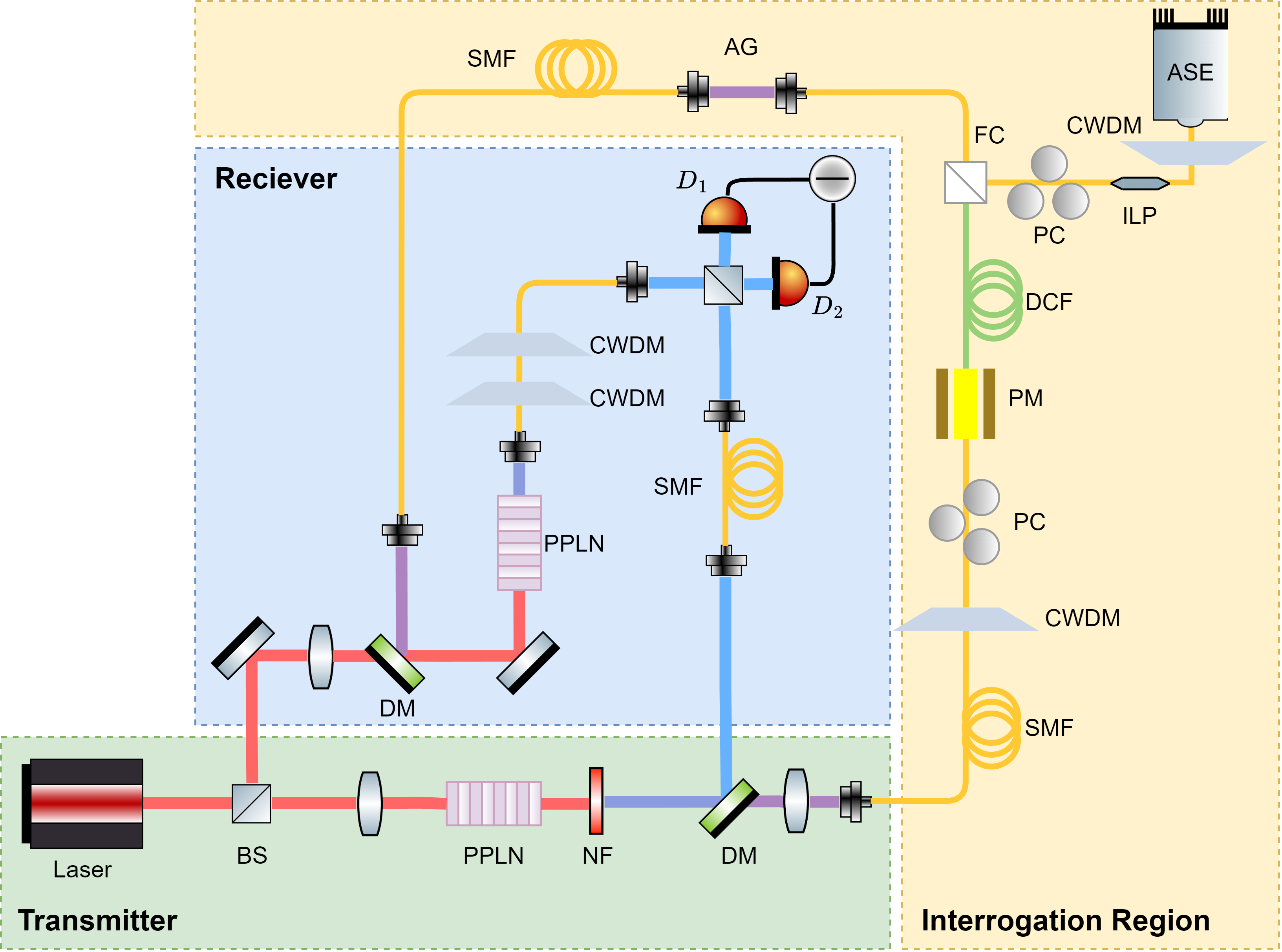}
        \label{subfig:exp_diagram}
    \end{subfigure}}
    \caption{ \small (a) Schematic of NP-QI, aiming to interrogate the absence or presence of a target; (b) Experimental diagram for NP-QI. BS, beamsplitter; DM, dichroic mirror; PPLN, periodically-poled lithium niobate; NF, notch filter; SMF, single-mode fiber; CWDM, coarse wavelength-division multiplexer; PC, polarization controller; PM, phase modulator; DCF, dispersion compensating fiber; ASE, amplifier spontaneous emission source; ILP, in-line polarizer; FC, fiber coupler; AG, air gap; D$_1$, D$_2$: detector. }
    \label{fig:concept}
\end{figure*}
\section{\label{sec:intro}Introduction}
Entanglement and superposition are key quantum resources to enable performance enhancements in sensing~\cite{quantum_sensing,photonic_q_sensing,EA_EM_sensing}, communication~\cite{EACOMM,covert_sensing}, and computing~\cite{shors_photonic}. Target detection is one such use case in which quantum resources can be leveraged to achieve superclassical performance~\cite{qi_story}. To detect a target, classical illumination (CI) interrogates the region of interest with coherent photonic modes emitted from a laser and subsequently takes a measurement on the reflected signal to determine whether it contains information about the target. In contrast, quantum illumination (QI) utilizes two entangled beams in two-mode squeezed-vacuum (TMSV) states, whereby one beam, the signal, interrogates the region of interest, while the other beam, the idler, is retained. The initial entanglement between the signal and idler facilitates QI's outperforming CI utilizing the same signal energy, quantified as a reduced error probability or enhanced signal-to-noise ratio (SNR)~\cite{qi_story}. The quantum advantage achieved by QI originates from the correlation between the entangled photonic modes that are stronger than any classical state can offer~\cite{qi_story}. A key feature of QI is its ability to retain performance benefits despite destruction of the initial entanglement between the probe and retained idler~\cite{ent_channel_survive}, rendering it particularly well-suited for target-detection systems operating in noisy and lossy environments.

To date, most experimental and theoretical advances in QI within the context of target detection~\cite{qi_gaussian,EE_Sensing} assumed equally likely target {absence} and {presence}~\cite{qi_story}. Further utilizing error probability as a performance metric \cite{qi_gaussian,Lloyd_qi,qi_story}, the analyses adopted by the prior QI studies were \textit{Bayesian}. However, radar theory prefers to sidestep Bayesian analysis~\cite{josab_npt}, as in target-detection environments of practical interest, it is often difficult to accurately assign prior probabilities to target absence and target presence, and appropriate costs to false-alarm (type-I) and miss (type-II) errors \cite{josab_npt}. Instead, radar theory opts for the Neyman-Pearson performance criterion for target detection, whereby the probability of detection $P_D = \textrm{Pr(present \textbar \  present)}$ is maximized subject to a constraint on the false-alarm probability $P_F = \textrm{Pr(present \textbar \ absent)}$, while the miss probability is defined as $P_M = \textrm{Pr(absent \textbar \ present)} = 1-P_D$. In this regard, a theoretical framework that provides an analytical expression for the optimum trade-off between the false-alarm and miss-probability error exponents in the asymptotic $(M \rightarrow \infty)$ limit using an $M$-copy quantum state has been established for QI~\cite{Spedalieri2014}. In addition, it has been shown that for a fixed false-alarm probability, QI's miss-probability exponent greatly outperforms CI's miss-probability exponent, while QI requires fewer trials to achieve this performance gain~\cite{Wilde2017}. Yet, to date an experimental protocol that exhibits a quantum advantage in detection schemes with unequal prior probabilities has not been demonstrated. 

Other QI-like experiments have been performed in the $N_B \ll 1$ regime \cite{Q_standoff,lopaeva2013experimental}, where $N_B$ is the number of background noise photons per mode. These protocols, however, do not hinge on entanglement and are not strictly applicable to radar systems, where $N_B \gg 1$ naturally occurs. On the other hand, more recent experiments, albeit not directly applicable to target detection as the original QI protocol addressed, have reported a 2.4 dB SNR advantage in entanglement-enhanced covert sensing~\cite{covert_sensing} and entanglement-assisted communication beyond the ultimate classical capacity~\cite{EACOMM}. From the perspective of target-detection systems, the extension of the QI platform into the unknown \textit{a priori} probability regime is important to bridge the gap with radar theory that prefers the {Neyman-Pearson} criterion. In radar applications, $P_F$ is often quite low, on the order of $10^{-4}$ to $ 10^{-8}$, because the cost of a false alarm can be extremely high \cite{Richards2022}. As such, lifting the known \textit{a priori} detection probability constraint will push it much closer to practical quantum-enhanced target-detection systems. Other methods of detection that do not rely on known \textit{a priori} detection probabilities exist \cite{Do2019,Richards2022}, however these detection protocols often involve several signal processing steps such as amplification and filtering that would serve to destroy and further degrade the initial entanglement purity generated by the source \cite{uwave_radar_prog}. 

In this work, we develop a Neyman-Pearson QI (NP-QI) system for target detection that exploits entanglement in a noisy and lossy environment to enhance the performance of target detection subject to unknown \textit{a priori} probabilities for the target being present and absent. Our experiment reports a quantum advantage obtained by leveraging QI in the optical domain, utilizing, for the first time, the Neyman-Pearson criterion as a performance metric in an unequal prior probabilities setting. The performance enhancement afforded by NP-QI is two-fold. For a fixed false-alarm probability $P_F$--which is often decided based on assigning appropriate costs to a false alarm--our NP-QI protocol enables a higher detection probability $P_D$. Conversely, at a fixed detection probability $P_D$, our NP-QI system enables a lower false-alarm probability $P_F$. Our work has important implications for target-detection systems, where the cost of a false-alarm that is not filtered appropriately by higher-level logic can be extremely high~\cite{Richards2022}. This result, along with the recent advancement in microwave QI reported by Assouly \textit{et al.} \cite{q_adv_uwave}, serves to motivate further investigation of QI in the microwave regime applied in the arbitrary prior probabilities space. \\

\section{\label{sec:results}Results }

\subsection{\label{sec:protocol}Target-Detection Protocol}

The entanglement source generates $M\gg 1$ independent, identically distributed (iid) signal-idler mode pairs with photon annihilation operators $\{\hat{a}^{(k)}_S,\hat{a}^{(k)}_I:1\leq k \leq M\}$. Each signal-idler mode pair is in a TMSV state, having an average photon number per mode $\braket{\hat{a}^{\dagger (k)}_S\hat{a}^{(k)}_S} = \braket{\hat{a}^{\dagger (k)}_I \hat{a}^{(k)}_I} = N_S$. The transmitter applies binary phase-shift keying (BPSK) on the signal modes to define the number of modes $M$ utilized by the receiver to produce a prediction, yielding $\{\hat{a}^{(k)}_S\}^M_{k=1} = \{(-1)^j\hat{a}^{(k)}_S\}^M_{k=1}$ for $j \in \{0,1\}$. The signal modes are sent to the interrogation region to probe the absence or presence of a target, while the idler modes are retained for the receiver. When the target is present, the phase-encoded {returned} signal modes are collected by the receiver after transmitting through a bosonic thermal-loss channel \cite{gaussian_qi} characterized by transmissivity $\kappa$, and an average noise photon number per mode $\braket{\hat{b}^{\dagger (k)}\hat{b}^{(k)}} = {N_B}/{(1-\kappa)}$. The transmissivity of the channel physically represents the reflectivity of a target. When the target is present, the returned signal modes are described by the following transform: $\{\hat{a}^{(k)}_R = \sqrt{\kappa} \hat{a}^{(k)}_S + \sqrt{1-\kappa}\hat{b}^{(k)}\}^M_{k=1}$. The received signal modes contain $N_B$ noise photons per mode on average. 

To determine the presence of a target, we construct a phase-conjugate receiver (PCR) to perform a joint measurement on the returned signal modes $\{\hat{a}^{(k)}_R\}^M_{k=1}$ and the retained idler modes $\{\hat{a}^{(k)}_I\}^M_{k=1}$~\cite{OPA}. In the PCR, phase-conjugate modes are produced through a parametric process with gain $G_A$: $\{\hat{a}^{(k)}_C = \sqrt{G_A}\hat{v}^{(k)} + \sqrt{G_A-1}\hat{a}^{\dagger (k)}_R\}^M_{k=1}$, where $\{\hat{v}^{(k)}\}^M_{k=1}$ are the vacuum field annihilation operators. The conjugate modes $\{\hat{a}^{(k)}_C\}^M_{k=1}$ then interfere with the retained idler modes $\{\hat{a}^{(k)}_I\}^M_{k=1}$ on a balanced beam splitter. This process produces the following modes on each respective output port of the beamsplitter.
\begin{equation}
    \left\{\hat{X}^{(k)} = \frac{1}{\sqrt{2}}\left(\hat{a}^{(k)}_C + \hat{a}^{(k)}_I\right)\right\}^M_{k=1}
\end{equation}
\begin{equation}
    \left\{\hat{Y}^{(k)} = \frac{1}{\sqrt{2}}\left(\hat{a}^{(k)}_C - \hat{a}^{(k)}_I\right)\right\}^M_{k=1}
\end{equation}
The receiver performs photon counting at each port, measuring $M$ modes simultaneously. In the limit $M \gg 1$, the detectors generate Gaussian variables $\{N_X,N_Y\}$. The difference photon number, defined as $N := N_X-N_Y$, reveals the \textit{phase-insensitive} cross correlations between the conjugate and idler modes $\{\braket{\hat{a}^{\dagger (k)}_C \hat{a}^{(k)}_I}\}^M_{k=1}$. Phase conjugation is required because the quantum signature of target presence is embedded in the \textit{phase-sensitive} cross correlations between the signal and idler modes of the form $\{\braket{\hat{a}^{(k)}_S \hat{a}^{(k)}_I}\}^M_{k=1}$, but phase-sensitive cross correlations cannot be revealed with second-order optical interference~\cite{qi_story}. The PCR thus serves to convert the phase-sensitive cross correlations into phase-insensitive cross correlations, which are detectable through second order optical interference~\cite{qi_story}. It is worth noting that the PCR's architecture is similar to that of the optical parametric amplifier (OPA)\cite{OPA,qi_story}. While both types of receivers can attain at most a $3 \ \textrm{dB}$ improvement in the SNR when compared to a CI platform of the same average energy\cite{qi_story,OPA}, PCR enjoys practical advantages due to common-mode noise cancellation in the balanced detection and higher tolerance to optical loss within the receiver~\cite{EACOMM}.

Neyman-Pearson target detection seeks to maximize the detection probability $P_D$ subject to a constraint on the false alarm probability $P_F$. To evaluate $(P_F,P_D)$ for a given dataset, we choose a threshold value $\beta$, such that a single-shot measurement produces a prediction from the set $\{H_0,H_1\}$. $H_0$ indicates that the NP-QI platform predicts target {absent} prediction, while $H_1$ indicates that target {present} is inferred.

We use the receiver operating characteristic (ROC) curve~\cite{josab_npt} to show that our NP-QI platform achieves a quantum advantage subject to the Neyman-Pearson criterion. The ROC is generated by collecting first and second moment statistics for each case of $h_0$ and $h_1$, where $h_0$ represents the ground truth of target being {absent}, and $h_1$ represents the known target {present} case. The threshold value $\beta$ is scanned, generating an associated $(P_F,P_D)$ data point for each $\beta$ value. The ROC is generated according to the following set of equations:
\begin{align}
    P_{F_i} &= Q_{h_0}\left(\frac{\beta}{\sigma}\right) \\
    P_{D_i} &= Q_{h_1}\left(\frac{\beta - \mu}{\sigma}\right),
\end{align}
where $Q_{h_j}(\cdot)$ is the $Q$-function, i.e., the tail probability of the normal distribution associated with $h_j$, and $(\mu,\sigma^2)$ are the first and second order moments of the target-{present} distribution. Note that $h_0$ and $h_1$ share the same variance $\sigma^2$ in the regime of $N_B \gg 1 \gg \kappa N_S$.

\begin{figure}[hbt!]
    \centering
    \includegraphics[width=\linewidth]{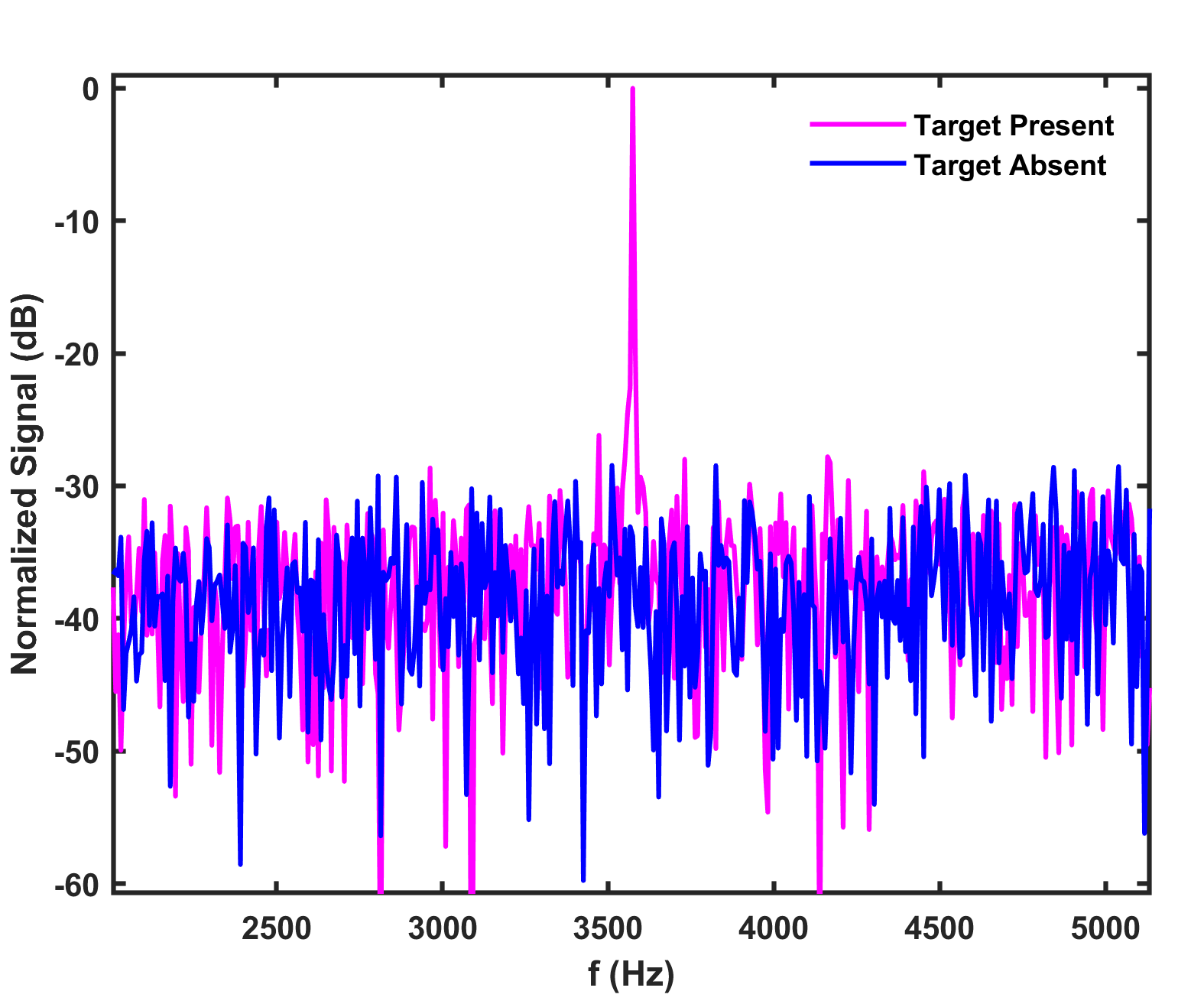}
    \caption{Frequency-domain trace, centered at the PM driving frequency $f_0=3571 \ \textrm{Hz}$, in both the target \textit{absent} (blue) and target \textit{present} (magenta) cases.}
    \label{fig:f_absent_present}
\end{figure}

We capture a frequency trace centered at the modulating frequency as displayed in Figure~\ref{fig:f_absent_present}, with the noise photon modes $\{N^{(k)}_B\}^M_{k=1}$ coupled into the signal channel, both in the target {absent} and {present} cases. The frequency-domain trace demonstrates our ability to resolve the fundamental frequency of the modulating signal, even in the presence of a bright noise background. It should be noted that unless the optical path lengths are matched to within the coherence length of the TMSV state ($\approx 150 \ \mu m $)\cite{halder2008high}, the modulating frequency peak is not present. The contrast between the signal peak and the noise floor at the fundamental modulating frequency $f_0$ is {35.8 dB}.

\begin{figure}[hbt!]
    \centering
    \includegraphics[width=\linewidth]{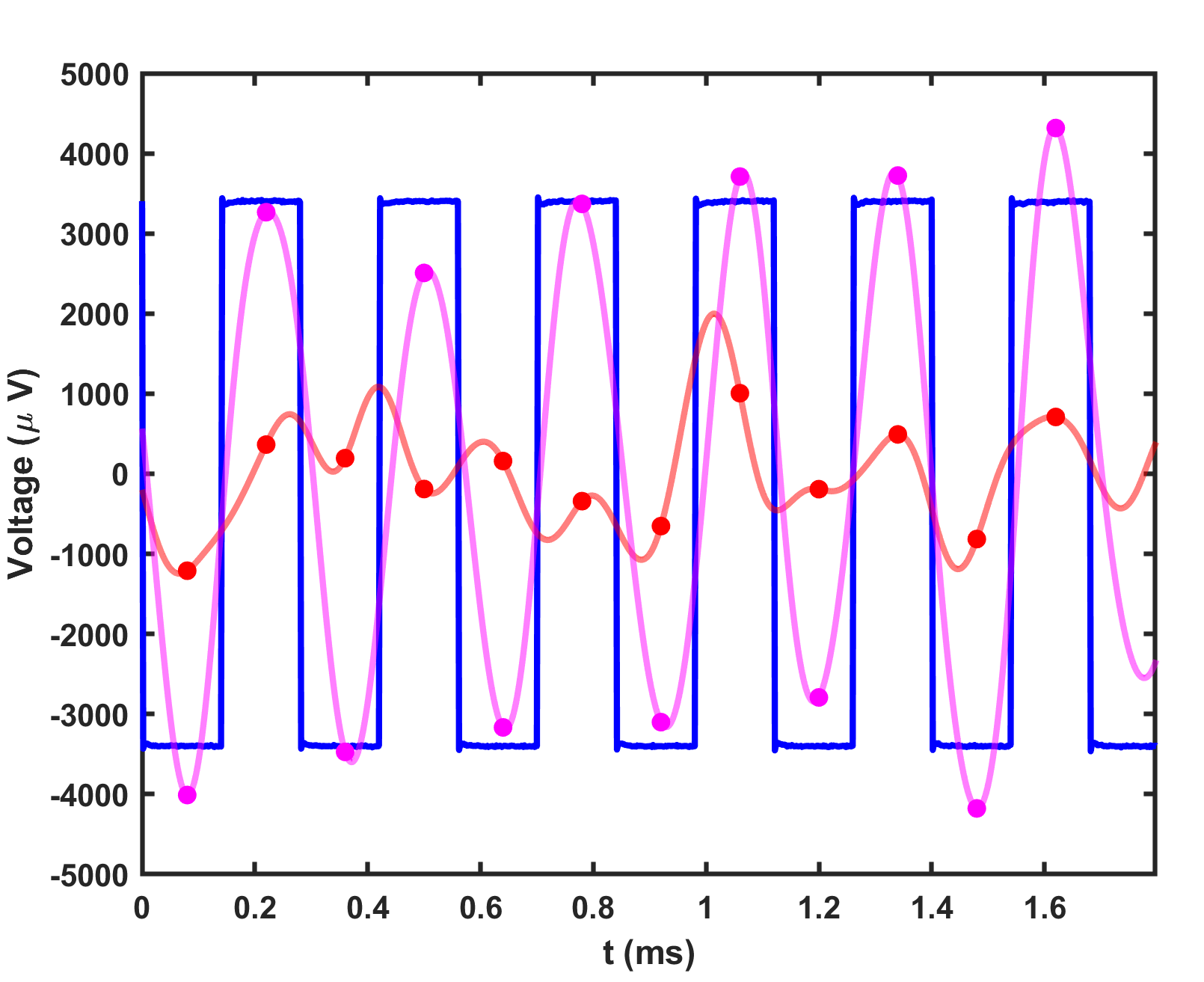}
    \caption{Time-domain traces of the PM driving signal, the target {present}, target {absent} data, and respective sampled datasets. The blue curve is the modulating signal applied to the electro-optic phase modulator on the signal arm. The magenta curve and points represent the time-domain signal and sampled data, respectively, in the target {present} case. The red curve and points represent the time-domain signal and sampled data, respectively, in the target {absent} case.}
    \label{fig:time_trace}
\end{figure}

We next plot in Fig.~\ref{fig:time_trace} a short time series of the output of the balanced homodyne detector in the $h_0$ and $h_1$ cases. The modulated signal is clearly distinguishable in the present case, in contrast to indistinguishable in the absent case; despite the fact that the per-mode noise brightness is much larger than that of the signal, i.e., $N_B \gg N_S$. 
The individually highlighted data points show the process of sampling the original data in post-processing, which results in a non-zero mean of the normal distribution in the presence of the target. The dataset derived from a selected phase of $0$ or $\pi$ is utilized to generate the ROC curve and assess the quantum advantage subject to the Neyman-Pearson criterion. 

\subsection{\label{sec:target_det_results}Target-Detection Measurement}
To show that our NP-QI platform affords a quantum advantage under the Neyman-Pearson criterion, we collect a time-series in the known target {absent} and {present} cases, a sample of which is shown in Figure~\ref{fig:time_trace}. Fundamentally, the ROC quantifies the ability to distinguish between two Gaussian distributions. Therefore, we sweep the decision threshold value $\beta \in \{-3\sigma_0,\mu_1+3\sigma_1\}$, where subscripts $0$, $1$ correspond to {absent} and {present}, respectively. For each $\beta$ value, the tail integral of each distribution is estimated to generate the corresponding $(P_F,P_D)$ data point. Under the known conditions $(h_0,h_1)$, corresponding to target {absent} and {present}, respectively, a probability density function $(\chi_0(v),\chi_1(v))$ is associated with the target {absent} and {present} conditions:
\begin{align}
    P_F = \int_{\beta}^{\gamma_0} dv \chi_0(v) \\
    P_D = \int_{\beta}^{\gamma_1} dv\chi_1(v), 
\end{align}
The integration is performed over binned voltage values generated from the time-series dataset, and $\gamma_i \  \forall \ i\in \{0,1\}$ corresponds to the maximum measured voltage for $(h_0,h_1)$, respectively. \\

\begin{figure*}[ht]
    \fbox{
    \begin{subfigure}{0.47\textwidth}
        \caption{\footnotesize}
        \includegraphics[width=\linewidth]{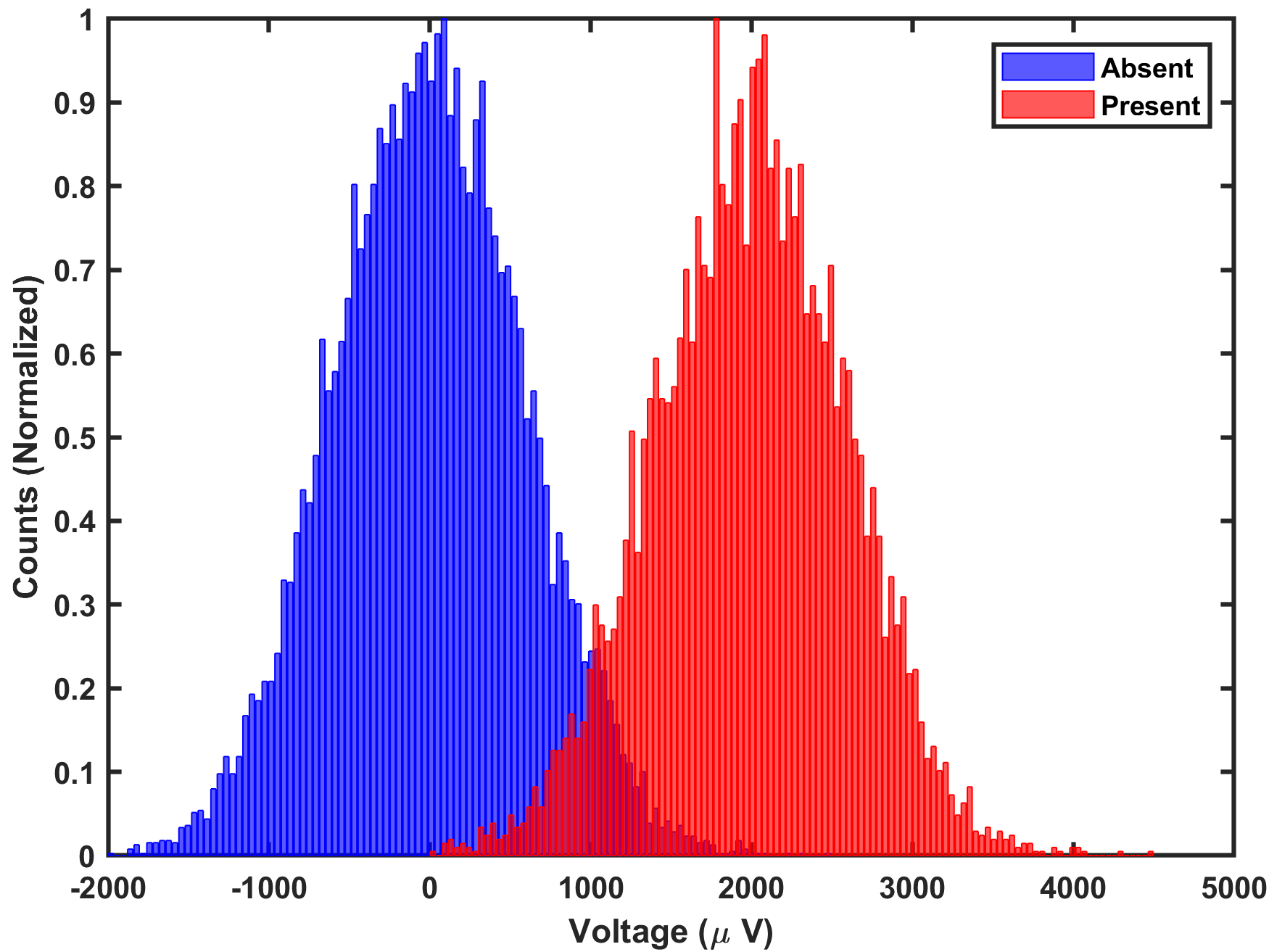}
        \label{subfig:classical_hist_M}
    \end{subfigure}\hfill
    
    \begin{subfigure}{0.47\textwidth}
        \caption{\footnotesize}
        \includegraphics[width=\linewidth]{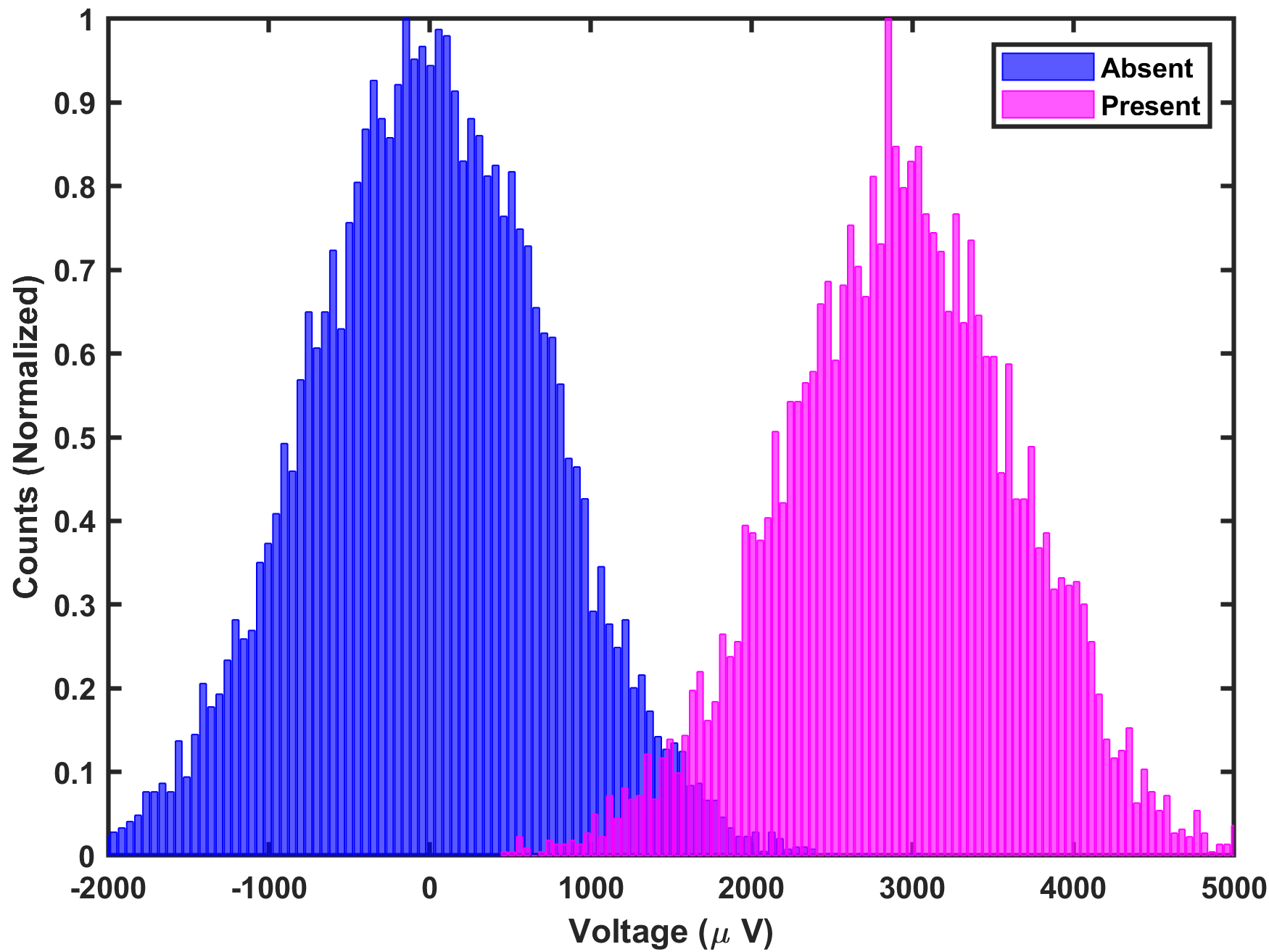}
        \label{subfig:quantum_hist_M}
    \end{subfigure}}
    \caption{\small Histograms for both target {absent} and {present} cases, illustrating the difference between the NP-CI and NP-QI platforms; (a) Simulated histograms for NP-CI for the target {present} (red) and {absent} (blue) cases; (b) Histograms from the NP-QI experimental platform, for the target {present} (magenta) and \textit{absent} (blue) cases.} 
    \label{fig:main_hist}
\end{figure*}

Signal-idler mode pairs produced by the entanglement source are iid~\cite{qi_story}, indicating that in the limit of $M\gg 1$ the photon number expectation value corresponding to detection $N=N_X-N_Y$ is Gaussian, which is completely characterized by its first and second order moments \cite{QI_supp_matl}. To derive the probability-density functions, we examine the histograms of the measurement data for the $h_0$ and $h_1$ cases. Figure~\ref{subfig:quantum_hist_M} depicts the histograms of the experimental data for the target {absent} and {present} cases, showing that the target {present} distribution is accurately modeled by a Gaussian distribution $\mathcal{N}(\mu,\sigma^2)$. As a comparison, Fig.~\ref{subfig:classical_hist_M} displays the histograms of the $h_0$ and $h_1$ cases for classical Neyman-Pearson classical illumination (NP-CI) for target detection operating with the same parameters of signal brightness, channel transmissivity, and noise power. It can be evidently observed that NP-QI enables a larger signal displacement while its variance remains comparable with that of the NP-CI. 

\begin{figure}[ht]
    \centering
    \includegraphics[width=\linewidth]{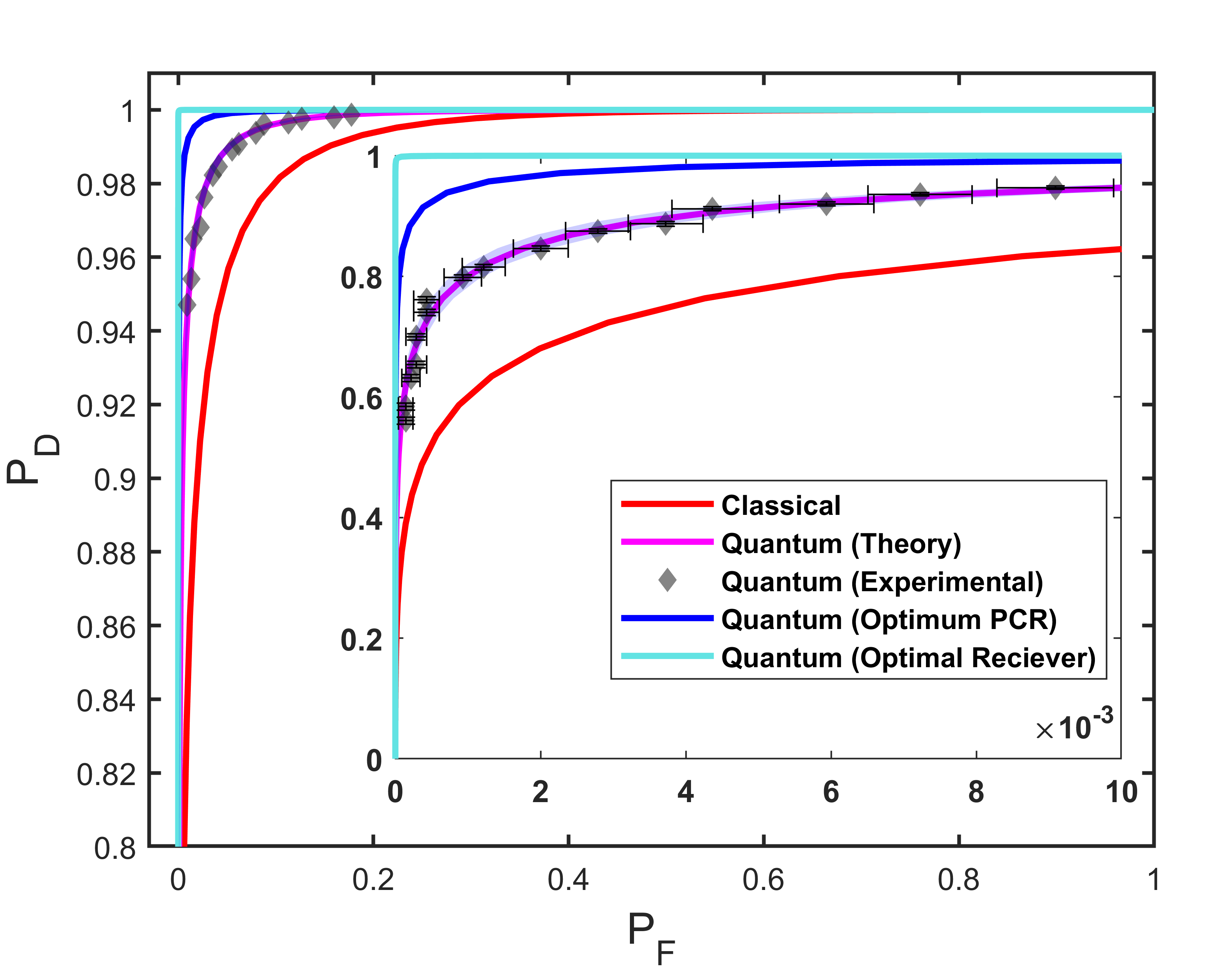} 
    \caption{Neyman-Pearson target-detection ROC curve. Red curve: ROC of the optimal NP-CI system~\cite{qi_gaussian}. Magenta curve: ROC of our NP-QI system, generated by fitting a normal distribution to the absent and present datasets and scanning the threshold value $\beta$. Diamond points: experimental data from the NP-QI system generated from the time series $h_0$ and $h_1$ datasets. Blue curve: the ultimate performance achieved by an ideal PCR receiver in NP-QI \cite{OPA}. Cyan curve: the ultimate performance achievable by QI based on an optimum receiver \cite{josab_npt,mixed_state_disc}. NP-CI and NP-QI operate with the same set of parameters of the signal brightness, channel transmissivity, and background noise level. Inset: zoom-in at small $P_F$ region. Parameters for all ROC curves: $M=10^{8.12}$, $N_S=6.8 \times 10^{-4}$, $\kappa = 0.086$, and $N_B=1.3\times 10^3$. Error bars for experimental data in the inset are calculated based on the statistical fluctuation of a binomial distribution, while they reside within the data points in the main plot. } 
    \label{fig:roc-main}
\end{figure}

We next generate an ROC curve based on a Gaussian fit to the absent $(h_0)$ and present $(h_1)$ cases, and superimpose this curve on $(P_F,P_D)$ data points measured directly from the time-series data. Figure \ref{fig:roc-main} illustrates the quantum advantage in the ROC achieved by our NP-QI platform, corresponding to a 1.48-dB enhancement in the SNR while the upper performance limit for a PCR is 3 dB derived from the receiver's quantum Chernoff bound~\cite{qi_story}. It is clear that our NP-QI system exhibits an advantage in that $P_D$ for the NP-QI system is strictly greater than that for the NP-CI system for all $P_F \in [0,1]$. The mean and variance predicted by our model accurately reproduces the ROC for coherent state homodyne detection from Zhuang \textit{et al.}~\cite{josab_npt}, which is a manifestation of the Bhattacharyya upper bound on the probability of error for CI \cite{qi_gaussian}. Our experimental setup is plagued by other noise contributions besides artificially added optical noise with $N_B$ photons per mode, such as electrical noise, optical phase noise, and mechanical vibrations. In contrast, the classical simulation does not include any additional noise sources in the model other than $N_B$ optical photons per mode. As such, NP-QI's quantum advantage subject to the Neyman-Pearson criterion is \textit{unconditional}, with potential to further enlarge the quantum advantage by suppressing the excess noise and improving the system stability. \\

The quantum advantage is most pronounced in the small $P_F$ region. This is a manifestation of the differences between the absent and present distributions in the NP-CI and NP-QI target detection. In both schemes, the target {absent} distributions are zero-mean, with a similar variance. However, there is a more significant difference between the target {present} distributions in the NP-CI and NP-QI cases. Although the standard deviation of the quantum target {present} distribution is larger, the difference between the mean value of the target {present} distributions exceeds the difference between the standard deviations. Therefore, in the small $P_F$ region, defined as $\beta > 2 \sigma_C$, $P_D$ for NP-QI is larger than that for NP-CI. In the large $P_F$ region, defined as $\beta < -2\sigma_C$, $P_{D}$s for NP-QI and NP-CI converge to a probability of $1$, as the entire $h_1$ distribution is sampled in both NP-QI and NP-CI cases. In addition, it is clear that the experimental data points match the ROC generated by sweeping $\beta$ over a Gaussian fit to the experimental data, as depicted in the magenta line in Fig.~\ref{fig:roc-main}. We observe that for small $P_F$, our ability to estimate $P_F$ becomes limited as the smallest $P_F$ that can be accurately estimated is set by the finite number of samples in the original dataset according to $\textrm{min}(\textrm{error rate}) \propto 10^{N-1}$ for $N$ samples \cite{proakis2008digital}. Thus, for an application where it is important to estimate small $P_F$ with high accuracy, a longer integration time would be required.

\section{\label{sec:Discussion}Discussion}
Our experiment demonstrates the optical NP-QI target-detection protocol's advantage over an equivalent classical system subject to the Neyman-Pearson criterion as bench-marked by an ROC curve. Ultimately, the promise of QI realized in the microwave regime is of more practical interest for the development of high impact technology \cite{qi_story,q_adv_uwave}. Although our experiment was performed in the optical domain, our work extends the body of knowledge related to QI protocols into the arbitrary prior probability space, paving the way for future microwave QI experiments that demonstrate a practical quantum advantage.
The ultimate performance enhancement limit that is physically achievable with a QI platform of the type presented in this work is a 6 dB reduction in the error-probability exponent, or equivalently, a 6-dB enhancement in the SNR, as formulated by Tan \textit{et al.}~\cite{qi_gaussian}, using tools developed by Pirandola and Lloyd \cite{q_chernoff}. Tan \textit{et al.}'s foundational theoretical analysis of a QI system that discriminates between $M$-copy quantum states made no mention as to the receiver hardware required to realize the maximum quantum advantage \cite{qi_gaussian,qi_story}. Guha and Erkamn proposed several receiver architectures to achieve a performance enhancement, however both were limited to a 3 dB enhancement \cite{OPA}. More recently, Zhuang \textit{et al.} proposed an architecture based on sum-frequency generation in tandem with feedforward, dubbed FF-SFG, which promises to extract the full 6-dB SNR enhancement from a QI platform \cite{mixed_state_disc,josab_npt}. However, an experimental realization of the FF-SFG receiver is yet to be reported. In this work, we utilize a PCR, whose quantum advantage is limited to a 3-dB enhancement in the SNR \cite{OPA}. We choose the PCR as opposed to the OPA type receiver, which was first demonstrated by Zhang \textit{et al.} with a 0.8 dB quantum advantage, because its performance is slightly improved compared to the OPA \cite{OPA}, due to cancellation of common mode noise in balanced dual detection \cite{OPA}.\\

Since the quantum advantage enabled by QI only prevails in high noise background brightness conditions \cite{qi_story}, where the number of noise photons per mode $N_B \gg 1$, artificially generated thermal background noise was produced in optical QI experiments to demonstrate its advantage. However, in the microwave domain, where most target-detection applications operate, these conditions are naturally satisfied \cite{qi_story}. The radar community's interest in microwave QI grew significantly after Barzanjeh \textit{et al.}'s proposal of a feasible microwave QI approach \cite{Barzanjeh2015}. Despite significant interest in implementing QI in the microwave domain, most attempts to achieve a quantum advantage have been unsuccessful due to the difficulty in preserving high-brightness entangled states once transmitted out of low-temperature, isolated environments \cite{uwave_radar_prog}, even though entangled fields with brightness comparable to optical SPDC sources have been generated in the microwave regime \cite{eichler2014quantum,flurin2012generating,sandbo2018generating}. Recently, Assouly \textit{et al.} demonstrated a $0.8 \ \textrm{dB}$ quantum advantage in a microwave radar implementation \cite{q_adv_uwave}, encouraging further exploration in this domain. Combining the strengths of the microwave and optical regimes offers new prospects for developing operational quantum technologies exhibiting more prominent advantages over their classical counterparts. In this regard, quantum transducers \cite{quics,Delaney2022,Lauk2020,Andrews2014,PhysRevX.12.021062} could play a pivotal role in accelerating the development of practical quantum technologies by enabling the conversion of quantum information between these spectral domains. In the context of QI, bidirectional microwave-to-optical quantum transducers will allow for the use of well-developed optical entanglement sources and quantum receivers, while transmitting microwave interrogating signals.

\section{\label{sec:conclusion}Conclusions}
Our QI system demonstrates an \textit{unconditional} quantum advantage over the optimal CI system in distinguishing between target {absent} and {present} cases subject to the Neyman-Pearson criterion. The achieved quantum advantage, albeit sub-optimal, persists across all $P_F$, suggesting that continued development of QI-based platforms would yield more pronounced quantum advantages and warrants further investigations and investment.

\begin{acknowledgments}
W. Ward acknowledges discussions with S. Liu. We gratefully acknowledge support from Office of Naval Research Award no. N00014-23-1-2296, National Science Foundation Award no. 2326746, US Department of Energy, Office of Science, National Quantum Information Science Research Centers, Superconducting Quantum Materials and Systems Center (SQMS), under contract no. DE-AC02-07CH11359, and University of Michigan.

\end{acknowledgments}

\appendix

\section{\label{sec:expp_setup}Experimental Scheme}
The experimental diagram for NP-QI is depicted in Figure~\ref{subfig:exp_diagram}. The entanglement source consists of a periodically-poled lithium niobate (PPLN) crystal, pumped by a laser at a wavelength of $780\,\textrm{nm}$, generating a broadband TMSV state centered at $1560\,\textrm{nm}$ through spontaneous parametric downconversion (SPDC). A notch filter rejects the pump at the PPLN output. The signal and idler modes are then separated by a dichroic mirror and are each coupled into a single-mode fiber (SMF). The optical bandwidth of the signal modes is set by a coarse wavelength division multiplexer (CWDM), which carves out a frequency band of $W=1.9 \ \textrm{THz}$ centered around $1590 \textrm{nm}$. 
 
In the target absent case $h_0$, the signal modes are disconnected from the channel, effectively setting the target reflectivity $\kappa=0$. In the target present case $h_1$, the number of modes $M$ used to make a decision about target absence or presence ($H_0 \ \textrm{or} \ H_1$), is defined by applying a $0 \ \textrm{or} \ \pi$ phase shift to the signal modes with an electro-optic modulator (EOM). The half period of the square wave signal $T$, together with the optical bandwidth $W$, defines $M = WT$. The signal modes are coupled into a dispersion-compensating fiber (DCF) to leverage the phenomenon known as nonlocal dispersion cancellation~\cite{nonlocal_disp}, which compensates for the dispersion incurred on both the signal and idler arms without introducing additional loss to the idler modes. An $L$-band unseeded amplified spontaneous emission source is used to generate thermal light due to its second-order coherence property and multimode photon statistics emulating a thermal state \cite{second_order_ase,pre_amp_ASE}. The thermal modes are mixed with the signal modes on the $10$ port of a $90:10$ fiber coupler. The combined signal and thermal noise modes are coupled to the receiver through a channel of transmissivity $\kappa$, which simulates a target in the interrogation region with reflectivity $\kappa$. \\
 
\par The PCR consists of a second PPLN nonlinear crystal pumped by the same $780\,\textrm{nm}$ light. The mixture consisting of signal and thermal modes passes through an air gap (AG), used to fine tune the relative delay between the signal and idler arms. The mixture is then coupled to free space through a collimator, and combined with the pump light with a dichroic mirror (DM). The mixture of modulated signal and thermal modes seeds a difference-frequency generation process with gain $G_A = 1+0.321 \times 10^{-3}$ at the PPLN crystal, producing phase-conjugate modes that are then coupled to an SMF through a collimator. The collected modes are sent through two cascaded $16 \textrm{nm}$-wide CWDMs centered at $1530 \textrm{nm}$ to reject the residue signal photons. The filtered phase-conjugate modes are coupled back to free space through a collimator and interfere with the idler modes that are coupled to free space through a collimator on a $50:50$ beamsplitter at a fringe visibility of $V\geq 98.5\%$ on both arms. The two output arms of the beamsplitter are detected by high quantum efficiency InGaAs photodiodes with a measured quantum efficiency $\eta_Q \geq 99\%$. The output electrical signal from the balanced detector is amplified using a trans-impedance amplifier, filtered by a low-pass filter with a bandwidth $BW = 0.70 \times (2 f_{\rm mod})$, where $f_{\rm mod}=3571 \ \textrm{Hz}$ is the driving frequency of the EOM. A servo loop is implemented to stabilize the relative phase between the pump, signal, and idler. The electrical signal is diverted to an electrical spectrum analyzer for frequency-domain measurements and an oscilloscope for time-domain measurements.

\section{\label{sec:theory}Theoretical Models}
\numberwithin{equation}{section}
\subsection{Classical-Illumination for Target Detection}
\begin{figure*}[ht]
    \centering
    \includegraphics[width=.8\linewidth]{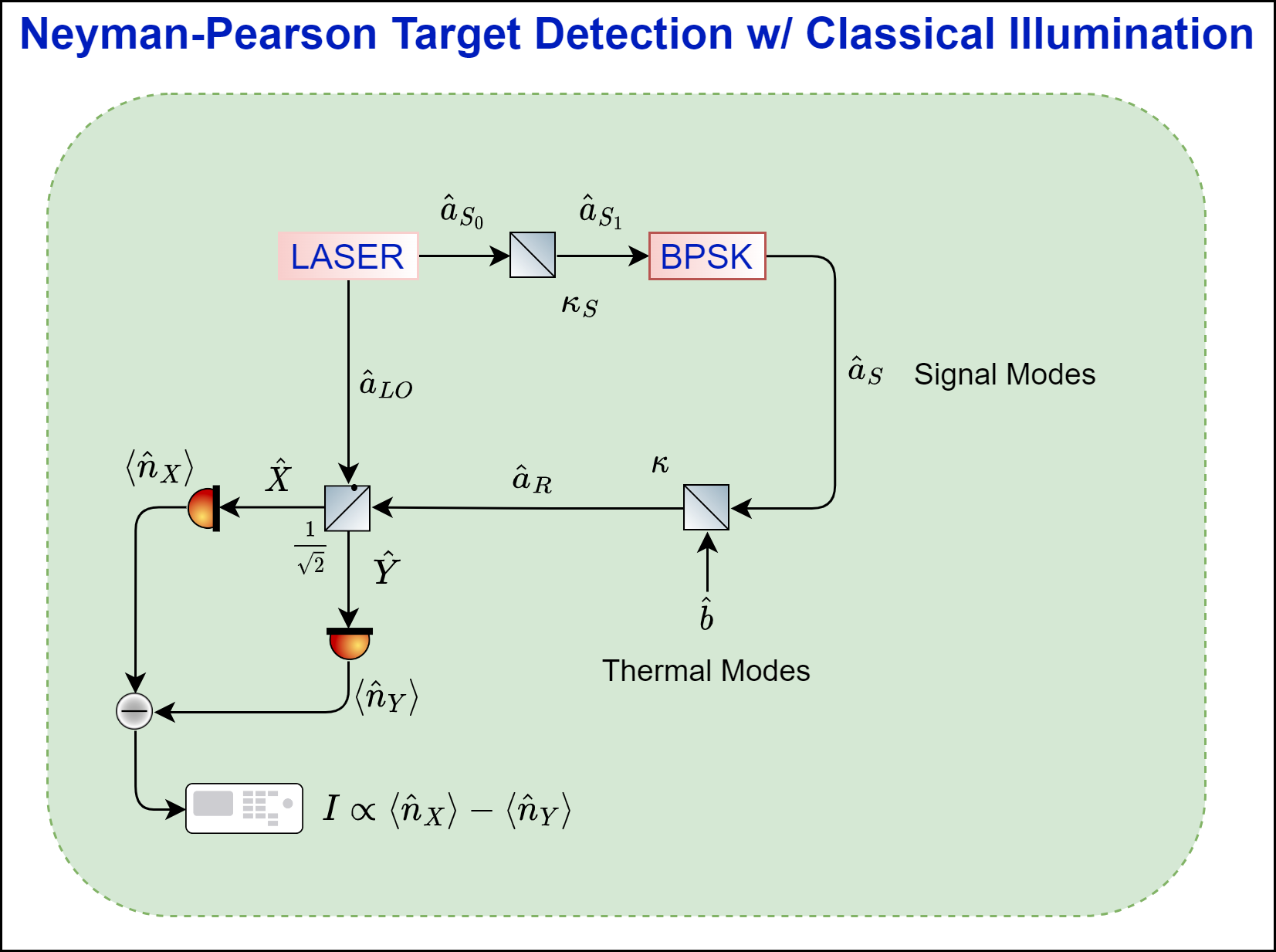}
    \caption{Schematic for Neyman-Pearson target detection with a classical light source}
    \label{fig:npt-classical}
\end{figure*}

We first formulate a model for a classical optical target-detection scheme described in Fig.~\ref{fig:npt-classical}, as a baseline to assess the quantum advantage that is experimentally realizable with NP-QI.

The signal modes $\{\hat{a}_S^{(k)}\} $ are sent through a channel of transmissivity $\kappa$, and mixed with thermal modes $\{\hat{b}^{(k)}\}$. The mixture is described by the following Bogoliubov transformation:
\begin{align}
    \hat{a}_R &= \sqrt{\kappa} \hat{a}_S + \sqrt{1-\kappa}\hat{b}.
\end{align}
Here, we work in a single-mode picture and drop the mode index $k$ for simplicity. Detection of target presence involves homodyne detection whereby the returned signal modes $\hat{a}_R$ are mixed with a local oscillator in a coherent state $\hat{a}_{\rm LO}$. We may now develop an expression for the homodyne signal using the appropriate quantum-optics formalism. We first describe the two output modes that are the result of mixing: $\{\hat{X},\hat{Y}\}$, then determine the homodyne signal for balanced detection:
\begin{align}
    \hat{X} &= \frac{1}{\sqrt{2}}(\hat{a}_R-\hat{a}_{\rm LO}) \\
    \hat{Y} &= \frac{1}{\sqrt{2}}( \hat{a}_R + \hat{a}_{\rm LO}). 
\end{align}
Now, we calculate the expected photon number at each detector, which follows from photodetection.
\begin{align}
    \braket{N_X} &= \braket{\hat{X}^\dagger \hat{X}}\\
    \braket{N_Y} &= \braket{\hat{Y}^\dagger \hat{Y}}.
\end{align}
The photocurrents from the two diodes are subtracted, yielding
\begin{align}
    \braket{N_Y} - \braket{N_X} &= \braket{\hat{a}^\dagger_R \hat{a}_{\rm LO} + \hat{a}^\dagger_{\rm LO} \hat{a}_R}.
\end{align}

At this juncture, we let the local oscillator be a coherent state with a known phase $\ket{\alpha_{\rm LO}} = |\alpha|e^{i \phi}$. Thus, the local oscillator operator is replaced with a complex number specified by its mean amplitude, yielding
\begin{align}
    \braket{N_Y}-\braket{N_X} &= |\alpha|\braket{e^{i \phi} \hat{a}^\dagger_R + e^{-i \phi} \hat{a}_R}
\end{align}
Since $\alpha$ is a scaling factor that equally applies to signal and noise, we let $\alpha = 1$ for simplicity. Now, we may transform the equation using Euler's relation to represent detection of the quadratures of the returned signal mode electric field:
\begin{align}
   \braket{N_Y}-\braket{N_X}
   =\sqrt{2} ( \textrm{cos}(\phi)\braket{\hat{Q}} +\textrm{sin}(\phi)\braket{\hat{P}}), \notag
\end{align}
where the quadratures $\hat{Q},\hat{P}$ are defined as follows:
\begin{align}
    \hat{Q} &:= \frac{1}{2}\left(\hat{a}^\dagger_R + \hat{a}_R\right) \\
    \hat{P} &:= \frac{1}{2i}\left(\hat{a}^\dagger_R - \hat{a}_R\right)
\end{align}
Now, we can set the LO phase $\phi=0$ to detect the $\hat{Q}$ quadrature. In taking this step, our homodyne signal becomes:
\begin{align}
    \braket{N_Y} - \braket{N_X} &= 2 \braket{\hat{Q}}
\end{align}
Now, we calculate the expectation value $\braket{\hat{N}_Y - \hat{N}_X}$ and the uncertainty (standard deviation) $\Delta (\hat{N}_Y-\hat{N}_X)$ of the observable produced by homodyne detection. The expectation value reads
\begin{align}
    &\braket{\hat{N}_Y-\hat{N}_X} = 2 \braket{\hat{Q}} \\
    &= \braket{\hat{a}^\dagger_R + \hat{a}_R}\notag \\
    &= \braket{\sqrt{\kappa}(\hat{a}^\dagger_S + \hat{a}_S) + \sqrt{1-\kappa}(\hat{b}^\dagger + \hat{b})}.\notag
\end{align} 
The expectation value of a thermal state is $\braket{\hat{b}} = 0$, while the expectation value of a coherent state is $\braket{\hat{a}_S} = \alpha_S$. We let the mean photon number of the $\hat a_S$ mode be $N_S = |\alpha_S|^2$ and apply these results to the expectation value:
\begin{align}
     &\braket{\hat{N}_Y-\hat{N}_X} = 2 \braket{\hat{Q}} \notag \\
     &= \sqrt{\kappa}(\alpha_S + \alpha^*_S) = 2\sqrt{\kappa}\textrm{Re}(\alpha_S) = 2\sqrt{\kappa N_S}
\end{align} 
Following a similar formalism, we can show
\begin{align}
    \Delta(\hat{N}_Y-\hat{N}_X) = \sqrt{2N_B+1}.
\end{align}

Now, since our target-detection task involves discriminating between target absence, whereby photon collection statistics generate a zero-mean Gaussian, and target presence whereby photon collection statistics generate a finite mean Gaussian, we may quantify the signal-to-noise ratio (SNR) in the classical case.
\begin{align}
    {\rm SNR}_{C_k} =\left(\frac{\mu_{C_k}}{\sigma_{C_k}}\right)^2,
\end{align}
where
\begin{align}
    \mu_{C_k} &=2\sqrt{\kappa N_S} \\
    \sigma_{C_k} &= \sqrt{2N_B +1}.
\end{align}
Thus, the SNR for a single signal-idler mode pair is derived as
\begin{align}
    {\rm SNR}_{C_k} &= \frac{4 \kappa N_S}{2N_B+1} \notag\\ 
\end{align}
With $M$ modes, the SNR simply scales up linearly:
\begin{align}
    {\rm SNR}_C = \frac{4\kappa N_S M}{2 N_B +1}.\notag \\
\end{align}
In the limit of $N_B \gg 1$, 
\begin{align}
\label{eq:SNR_C}
     {\rm SNR}_C = \frac{2\kappa N_S M}{N_B}. \notag
\end{align}
With this definition, we may now redefine the first and second moments $(\mu_C,\sigma_C)$ for the classical case as follows, where the mean and variance are multiplied by $M$, the number of modes, because our preceding analysis was executed in a single-mode basis. 
\begin{align}
    \mu_C &= \sqrt{2\kappa N_S}M \\
    \sigma_C &= \sqrt{N_B M}
\end{align}
in the limit of $N_B \gg 1$.

\subsection{Quantum-Illumination for Target Detection}
A simplified diagram of the quantum target-detection model is shown below in Fig.~\ref{fig:npt-q-simplified}. The signal photon number per mode and the noise photon number per mode are defined in accordance with the classical model to facilitate an equitable comparison of the two protocols.
\begin{figure*}[ht]
    \centering
    \includegraphics[width=\linewidth]{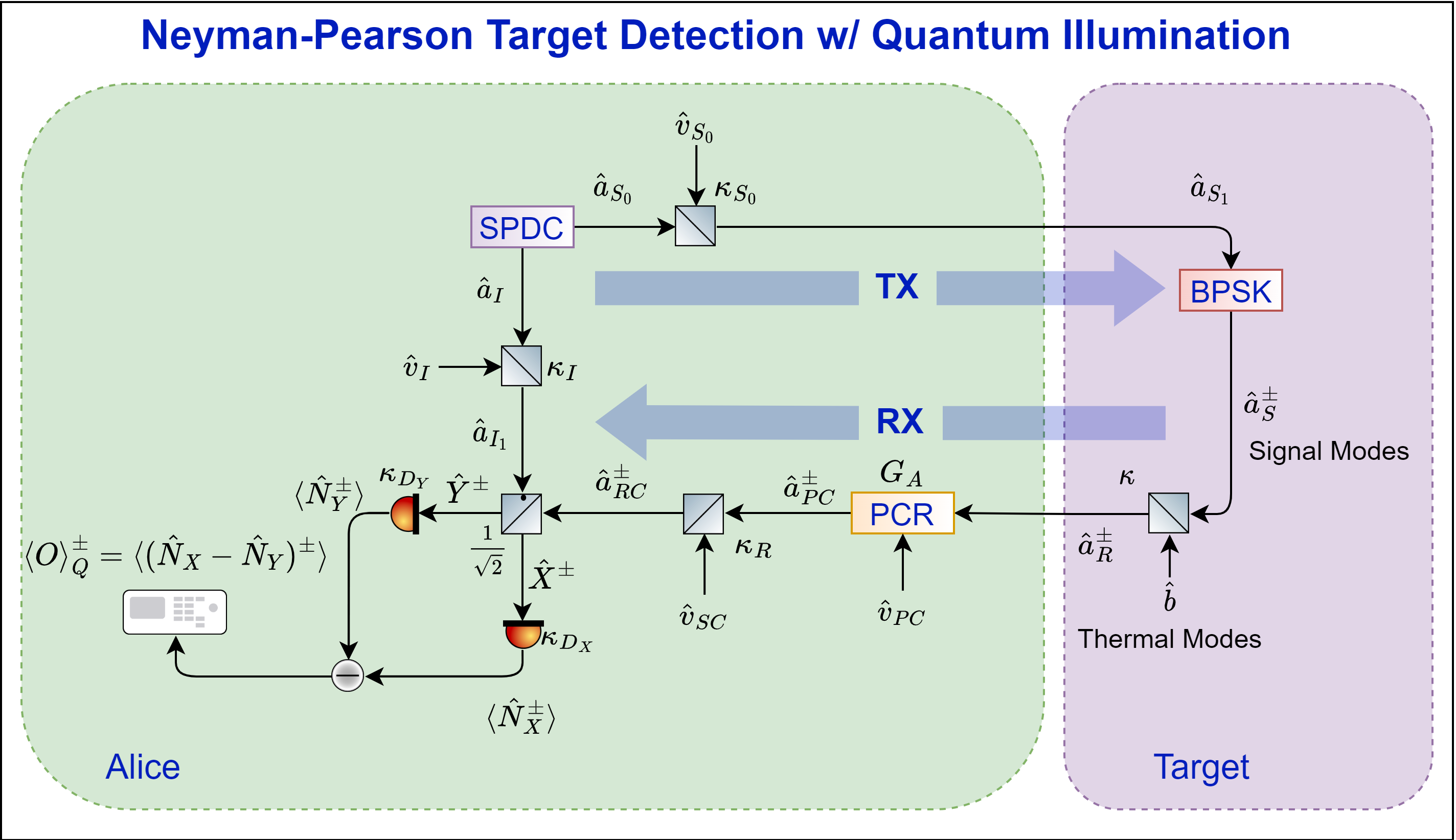}
    \caption{NP-QI target detection with an entanglement source and quantum receiver.}
    \label{fig:npt-q-simplified}
\end{figure*}
Now, in accordance with the classical model, we define the signal modes $\hat{a}^\pm_S$ with photon number per mode $N_S$, and the thermal noise modes $\hat{b}$ with photon number per mode $N_B/(1-\kappa)$, which are combined and transmitted signal through a channel of transmissivity $\kappa$. We now complete a single mode analysis to determine the first two moments of the distribution produced by detection of the relevant observable produced by homodyne detection, defined as follows.
\begin{align}
    \braket{O}^\pm_Q &= \braket{(\hat{N}_X -\hat{N}_Y)^\pm}
\end{align}
First, we propagate the relevant modes through the illumination system. Then, we determine the first and second moments for each modulation case, which is mapped as follows: $\theta = (0,\pi) \rightarrow (+,-)$. The modes that interact with the receiver are defined as $\hat{a}^\pm_R$, according to the following transformation:
\begin{align}
    \hat{a}^\pm_R = \sqrt{\kappa} \hat{a}^\pm_S + \sqrt{1-\kappa}\hat{b}.
\end{align}
The received modes are phase conjugated by a PCR with gain $G_A$ according to the following Bogoliubov transformation:
\begin{align}
    \hat{a}^\pm_{\rm PC} = \sqrt{G_A} \hat{v}_{\rm PC} + \sqrt{G_A -1}\left(\hat{a}^\pm_R\right)^\dagger.
\end{align}
The phase-conjugate modes are then sent to a balanced homodyne detection apparatus through a channel with transmissivity $\kappa_R$. 
\begin{align}
    \hat{a}^\pm_{\rm RC} &= \sqrt{\kappa_R} \hat{a}^\pm_{\rm PC} + \sqrt{1-\kappa_R} \hat{v}_{\rm SC}.
\end{align}
We can ignore the vacuum modes because they will evaluate to zero when we evaluate relevant expectation values. We may now define the modes that result from mixing the conjugate signal and the retained idler modes:
\begin{align}
    \hat{Y}^\pm &= \frac{1}{\sqrt{2}}(\hat{a}^\pm_{\rm RC} - \hat{a}_{I_1} )\\
    \hat{X}^\pm &= \frac{1}{\sqrt{2}}(\hat{a}^\pm_{\rm RC} + \hat{a}_{I_1} ).
\end{align}
Here, $\hat{a}_{I_1} = \sqrt{\eta_I} \hat{a}_{I} + \sqrt{1-\eta_I} \hat v_I$, where $\eta_I$ is the transmissivity of the idler storage channel and $\hat v_I$ is a vacuum mode. 

Following detection, the expected photon number at each photodetector is defined as follows:
\begin{align}
    \left<\hat{N}^\pm_Y\right> &= \left<\left(\hat{Y}^\pm\right)^\dagger \hat{Y}^\pm\right> \\
    \left<\hat{N}^\pm_X\right> &= \left<\left(\hat{X}^\pm\right)^\dagger \hat{X}^\pm\right>
\end{align}
At this point, we will treat the $(+,-)$ cases separately, because the statistics are required for both cases to derive the signal-to-noise ratio for the quantum-enhanced illumination system.
\subsubsection{Statistics for $\theta=0$ Phase Modulation}
First, we treat the $(+)\rightarrow \theta=0$ case, where no phase modulation is applied. We drop the $+$ superscript within this section, recognizing that we are only working with the $+$ phase modulation case.
\begin{align}
    \braket{\hat{N}_Y} &= \braket{\hat{Y}^\dagger \hat{Y}} = \frac{1}{2}\braket{(\hat{a}^\dagger_{\rm RC} - \hat{a}^\dagger_{I_1})(\hat{a}_{\rm RC} - \hat{a}_{I_1})} \\
    \braket{\hat{N}_X} &= \braket{\hat{X}^\dagger \hat{X}} = \frac{1}{2}\braket{(\hat{a}^\dagger_{\rm RC} + \hat{a}^\dagger_{I_1})(\hat{a}_{\rm RC} + \hat{a}_{I_1})}\\
    \braket{\hat{O}}^+_Q &= \braket{\hat{N}_X - \hat{N}_Y} = \braket{ \hat{a}^\dagger_{\rm RC} \hat{a}_{I_1} + \hat{a}^\dagger_{I_1} \hat{a}_{\rm RC} }
\end{align}
Now, we may evaluate the mean and variance of the $+$ case homodyne observable $\hat{O}^+_Q$.
We also utilize the fact that $\hat{b}$ represents a thermal state and obtain
\begin{align}
    \braket{\hat{O}}^+_Q &= 2\Lambda \sqrt{\kappa N_S (N_S+1)}
\end{align}
We see that the quantum observable has revealed the phase-sensitive cross correlations between signal and idler beams, as expected. These correlations contain the quantum-enhanced signature of target presence. The next step is to derive the standard deviation of the $+$ observable $\Delta \hat{O}^+_Q$.
\begin{align}
    \Delta\hat{O}^+_Q &= \sqrt{\left<\left(\hat{O}^+_Q\right)^2\right> - \left(\left<\hat{O}^+_Q\right>\right)^2} 
\end{align}
In the parameter region of interest, one can show that
\begin{align}
    \Delta\hat{O}^+_Q \approx \sqrt{\kappa_R(G_A-1)(N_B+1)}
\end{align}

The operating regime of QI is restricted to: $(N_B \gg 1 \gg N_S)$. Therefore, only several terms survive in the standard deviation, yielding
\begin{align}
\Delta\hat{O}^+_Q \approx \sqrt{\kappa_R(G_A-1)N_B}
\end{align}
Multiplying ($\mu^+_Q,{\sigma^+_Q}^2$) by the number of modes $M$ consumed during detection of $0$ phase modulation and eliminate the $\kappa_R$ and $G_A-1$ terms that are present in both $\mu^+_Q$ and ${\sigma^+_Q}^2$, we obtain
\begin{align}
    \mu^+_{Q_k} &= 2\sqrt{\kappa N_S (N_S+1)} \notag\\
    \sigma^+_{Q_k} &= \sqrt{N_B} \notag\\
    \mu^+_Q &= 2\sqrt{\kappa \kappa_I N_S(N_S+1)}M \notag \\
    \sigma^+_Q &= \sqrt{N_BM}
\end{align}

\subsubsection{Statistics for $ \theta=\pi$ Phase Modulation}
The derivation up to a certain point of the $\theta=\pi$ case modulation is equivalent to the $\theta=0$ phase modulation case, so we will use the appropriate equation above as a starting point. We use $\Lambda$ as defined in the previous section and utilize the fact the mode operator $\hat{b}$ represents a thermal state.
\begin{align}
    \braket{\hat{O}}^-_Q &= -2\Lambda \sqrt{\kappa N_S(N_S+1)} 
\end{align}
The uncertainty of the $\theta=\pi$ phase modulation case is equivalent to the $\theta=0$ phase modulation case. This result manifests from following through with the algebra as in the previous section. However, from a qualitative point of view this result is expected because a different phase on one of the arms of an interferometer would not be expected to change the variance of a homodyne measurement.
\begin{align}
\Delta \hat{O}^-_Q &= \sqrt{\left<\left({\hat{O}^-_Q}\right)^2\right> - \left(\left<\hat{O}^-_Q\right>\right)^2} \notag \\
&\approx \sqrt{\kappa_R(G_A-1)N_B}
\end{align}
We recognize that when we determine the SNR, the term $\Lambda$ will cancel. We also move the factor of $2$ in the quantity $\mu^-_Q$ into the uncertainty term $\sigma^-_Q$ as we will eventually examine their ratio. Therefore, we define the following quantities. We remind the reader that our analysis was performed in the single mode basis. In order to move to the multi-mode case, we multiply the mean and variance by the number of modes $M$ consumed by detection of the $\theta=\pi$ case and eliminate the $\kappa_R$ and $G_A-1$ terms that are present in both $\mu^+_Q$ and ${\sigma^+_Q}^2$, we obtain
\begin{align} 
    \mu^-_{Q_k} &=  -2\sqrt{\kappa N_S (N_S+1)} \notag\\
    \sigma^-_{Q_k} &= \sqrt{N_B} \notag\\
    \mu^-_Q &= -2\sqrt{\kappa N_S (N_S+1)}M \notag\\
    \sigma^-_Q &=\sqrt{N_B M}.
\end{align} \\

\subsection{Comparison of QI and CI SNRs}
Now, we may determine ${\rm SNR}_Q$, the quantum SNR. The definition of the SNR is altered slightly due to the nature of our target-detection technique. With the quantum system, our task is to distinguish between two Gaussian random variables produced by $\theta = (0,\pi) \rightarrow (+,-)$ phase modulation. The two distributions have the same variance. Therefore, we may define ${\rm SNR}_Q$ as follows:
\begin{align}
    {\rm SNR}_Q &= \frac{|\mu^\pm_Q|^2}{{\sigma^\pm_Q}^2} \notag \\
    {\rm SNR}_Q &= \frac{4M \kappa\kappa_I N_S (N_S+1)}{N_B} 
\end{align} 
Since QI operates in the $N_S \ll 1$ regime, we may discard terms ${N_S}^2$ because ${N_S}^2 \ll N_S$ when $N_S \ll 1$. The quantum SNR in this parameter region becomes the following.
\begin{align}
    {\rm SNR}_Q &= \frac{4 M \kappa\kappa_I N_S}{N_B}
\end{align}
To model extra system imperfections such as sub-optimal phase locking, imperfect dispersion compensation, and sub-unity detector efficiency, we introduce a fitting parameter $\zeta$ in the quantum SNR and obtain:
\begin{align}
    {\rm SNR}_Q &= \frac{4 M \zeta\kappa\kappa_I N_S}{N_B}.
\end{align}
$\zeta$ can be extrapolated from experimental data. Comparing with Eq.~\ref{eq:SNR_C}, SNR$_Q$ is two times SNR$_C$ in an ideal scenario of no imperfections, i.e., $\kappa_I = \zeta = 1$.

\bibliography{references}

\end{document}